\documentclass{aa}
\usepackage[varg]{txfonts}
\usepackage[autostyle, english=british]{csquotes}
\usepackage{hyperref}
\usepackage{cleveref}
\usepackage{amsmath, amssymb, bbold, mathrsfs, nicefrac, dsfont, mathtools, booktabs}
\usepackage{pgf,siunitx}
\usepackage{soul}
\DeclareSIUnit{\muas}{\mu\mathrm{as}}
\DeclareSIUnit{\as}{\mathrm{as}}
\DeclareSIUnit{\arcmin}{\mathrm{arcmin}}
\DeclareSIUnit{\parsec}{pc}
\DeclareSIUnit\gaussunit{G}
\DeclareSIUnit\jy{Jy}
\graphicspath{{auto/}}

\newcommand{\iu}{{i\mkern1mu}}

\newcommand{\resolve}{\texttt{resolve}}
\newcommand{\clean}{\texttt{CLEAN}}

\newcommand{\loadplotpng}[1]{\includegraphics{auto/#1.png}}

\begin{document}

\title{Bayesian Imaging of Interferometric Data from Polarized Electromagnetic Signals}
\author{Philipp Arras\inst{\ref{mpa},\ref{tum}} \and Jakob Roth \inst{\ref{mpa}, \ref{tum}} \and Martin Reinecke\inst{\ref{mpa}} \and Richard A. Perley\inst{\ref{nrao}} \and Andrei Frolov\inst{\ref{sfu}} \and Rüdiger Westermann\inst{\ref{tum}} \and Torsten~A.~En{\ss}lin\inst{\ref{mpa},\ref{lmu},\ref{dza}}}
\institute{
Max-Planck Institut für Astrophysik, Karl-Schwarzschild-Str. 1, 85748 Garching, Germany\label{mpa}
\and
Ludwig-Maximilians-Universität München (LMU), Geschwister-Scholl-Platz 1, 80539 München, Germany\label{lmu}
\and
Technische Universität München (TUM), Boltzmannstr. 3, 85748 Garching, Germany\label{tum}
\and
Simon Fraser University, 8888 University Drive, Burnaby BC, Canada\label{sfu}
\and
National Radio Astronomy Observatory, P.O. Box O, Socorro, NM 87801, USA\label{nrao}
\and
Deutsches Zentrum f{\"u}r Astrophysik, Postplatz 1, 02826 G{\"o}rlitz, Germany\label{dza}
}
\date{Received <date>/ Accepted <date>}
\abstract{
  We present an imaging algorithm for polarimetric interferometric data from radio telescopes.
  It is based on Bayesian statistics and thereby able to provide uncertainties and to incorporate prior information such as positivity of the total emission (Stokes I) or consistency constraints (polarized fraction can only be between 0\% and 100\%).
  By comparing our results to the output of the de-facto standard algorithm called \clean{}, we show that these constraints paired with a consistent treatment of measurement uncertainties throughout the algorithm significantly improve image quality.
  In particular, our method reveals that depolarization canals in \clean{} images do not necessarily indicate a true absence of polarized emission, e.g., after frequency averaging, but can also stem from uncertainty in the polarization direction.
  This demonstrates that our Bayesian approach can distinguish between true depolarization and mere uncertainty, providing a more informative representation of polarization structures.
}
\maketitle


\section{Introduction}

Polarization imaging in radio astronomy plays a crucial role in understanding a wide variety of astrophysical phenomena ranging from planets, stars, over supernovae remnants, the interstellar medium of the Milky Way, to other galaxies, the intergalactic medium and even whole galaxy clusters.
Polarized radio emission typically results from synchrotron emission of relativistic particles gyrating in magnetic fields \citep{Ginzburg65,Legg68}.
These particles were initially accelerated at shock waves, magnetic reconnection events, and in plasma turbulence.
The plane of polarization of the electric fields of their synchrotron emission is perpendicular to the magnetic field that causes them to emit.
Thus, observations of polarized radio signals carry invaluable information about the physical mechanisms and structures of these celestial objects that remain elusive in total intensity observations \citep{Pacholczyk70}.
Mapping accurately the polarization structure of such radio sources like radio lobes of galaxies, supernova remnants, and others, therefore provides detailed insight into their inner workings.

Furthermore, the plane of polarization of radio synchrotron emission is rotated on its way to us due to the Faraday effect whenever it traverses a magnetized plasma.
This imprints a characteristic change in polarization as a function of wavelength onto the radiation and thereby provides detailed insights into the environments through which these radio waves have propagated, providing a way to map out the magnetic field structure between emitter and observer.
A remarkable technique to analyze polarized Faraday rotated radio emission and to some degree 3d information about magnetic fields is Faraday tomography \citep{Simard81,Dreher87,Brentjens05,2024A&A...692A.248G}.

Linearly polarized radio emission is also converted into circularly polarized light and vice versa due to the Faraday conversion effect, a slightly different interaction between light and magnetized plasma \citep{Jones77}.
This effect has a stronger dependence on the magnetic field strength and observing wavelength and is therefore mostly seen in compact radio sources \citep{1984MNRAS.208..409K,1984Ap&SS.100..227V,1990MNRAS.242..158B,1999ApJ...523L..29B,2000ApJ...530L..29F,2001ApJ...560L.123B,2002A&A...388.1106B,2002ApJ...573..485R,2003A&A...401..499E,2019ApJ...873...55B,2021Galax...9...58G,2022ApJ...931...25T}.
However, there is a realistic possibility to observe circularly polarized emission due to Faraday conversion in bright hotspots of radio galaxies and even from the entire Milky Way \citep{2017PhRvD..96d3021E,2019JCAP...01..035E}.

Thereby, polarization imaging plays a pivotal role in deriving insights into magnetized environments in the cosmos, e.g., radio sources and the magnetized ionized media they are embedded in.
It allows us to investigate the magnetic field orientation and strength in various celestial objects, providing an indispensable tool in astrophysics.
Faraday rotation measures from polarization imaging provide crucial insights into the magnetic fields present in interstellar and intergalactic media.

Imaging in radio astronomy, especially with interferometry, comes with many challenges.
The main difficulty is that the measurements are noisy and incomplete in terms of Fourier space coverage.
As a result, converting raw data into useful scientific information is nontrivial and requires robust methodologies and consistent data processing.
There are several approaches to tackle this problem, each with its own advantages and limitations.

The goal of any imaging method is to provide images of the sky that are as close to reality as possible in a certain sense.
For this the algorithms make use of the data, but might differ in how they do this, they exploit knowledge about the sky to a varying degree, differ in the numerical treatment of the imaging problem, and even have different objectives (being accurate with respect to a specific error norms, being computationally inexpensive, being accepted by the community, or the like).
As we want to show the benefits of using a Bayesian framework for polarization imaging, we emphasize here the aspects of algorithms that are improved (e.g.\ image fidelity and interpretability, reflection of physical constraints, uncertainty characterization) over those that are not improved (e.g.\ processing speed).

Historically, imaging of polarized emission has been achieved through the deployment of a maximum likelihood approach integrated with an effective regularization, typically provided by the \clean{} imaging algorithm.
The basic incarnation of this method involves the search and subsequent subtraction of point sources within the imaged field until only background noise remains.
This method is versatile and can be applied to both Stokes~$I$ imaging and comprehensive polarization imaging.
In the context of polarization imaging, the algorithm independently executes greedy peak searches on the Stokes~$I$, $Q$, $U$, and $V$ image components.
This traditional approach has been supplemented and enhanced through a variety of methods.
For instance, \citet{pratley16} proposed to improve correlation amongst the Stokes components by seeking peak intensities in total polarized emission $P$ as represented by the equation $P^2=Q^{2} + U^{2} + V^{2}$, thereby establishing implicit correlation among these components.
A robust example of potent polarization \clean{} reconstruction is exemplified in the work done by \citet{sebokolodi2020wideband}, which we take for comparison by using the same dataset.

The \clean{} imaging algorithm, like any method, presents a combination of advantages and potential shortcomings.
On the positive side, \clean{} distinguishes itself by its high computational speed, making it a very efficient tool that is widely recognized and used within the astronomical community.
Its simplicity makes it easy to understand, allowing for straightforward debugging during image reconstruction.
Moreover, there exist multiple freely available implementations of the algorithm, such as CASA and WSClean, making it accessible for a broad range of applications.
Conversely, \clean{} has certain limitations that can impact its performance and applicability.
One such limitation is its naivety regarding basic constraints, such as the positivity of Stokes I emission and the consistency constraints on polarization emission ($I \ge P$).
Ignoring this can lead to potential inaccuracies in image reconstructions.
The algorithm also underperforms in extremely low-noise regimes, which can limit its utility in certain observational scenarios.
Notably, \clean{} does not inherently provide uncertainty information on its image outputs.
This lack of error quantification makes it challenging to rigorously assess the reliability of the derived scientific insights, which could have substantial implications for interpretations and conclusions drawn from data.
So while \clean{} remains a valuable tool, these caveats require special care in polarisation imaging.

Alternative imaging methodologies, although not as mainstream as the \clean{} algorithm, also play significant roles within the radio astronomical community.
One such paradigm consists of maximum entropy methods as described in various studies \citep{Holdaway90,hamaker96,Coughlan2016,Chael16}.
These approaches allow for the inclusion of polarization and accommodate certain forms of constraints, such as that total intensity must exceed polarized intensity.
Compressed sensing algorithms further embody an important category of radio imaging techniques.
\citet{birdi18} illustrated the application of compressed sensing with full polarization, while adhering to polarization constraint $I\ge P$, on simulated Event Horizon Telescope (EHT) data.
This method was later augmented to incorporate direction-dependent calibration \citep{birdi19}.
Within the scope of Very Long Baseline Interferometry (VLBI) imaging \citet{akiyama17} advocated for an imaging methodology founded on total variation regularization (specifically, $l_{1}$ regularization), with the goal to minimize data residuals.
Implementation of this technique on simulated M87* EHT data was successful.
However, it should be noted that $l_{1}$ regularization, while effective, is not a consistent regularization in the strict sense, since it depends on the chosen space discretization scheme.
The appropriate choice of imaging protocol should therefore be guided by the specific context and requirements of the observational data.

Another central set of imaging techniques is based on Bayesian statistics and information field theory \citep{ensslin09,ensslin11,ensslin18}.
Specifically, the \resolve{} algorithm \citep{junklewitz2016resolve,arras2021comparison} is an example of such.
The characteristic strengths of this approach contrast with those of the aforementioned methods.
\resolve{} explicitly states its theoretical assumptions, which improves the understanding and predictability of its performance.
Further, it is versatile, capable of addressing a wide range of needs, including traditional interferometric imaging, direction-independent calibration \citep{arras2019unified}, direction-dependent calibration \citep{roth2023bayesian}, and VLBI, wherein closure quantities are employed in the likelihood process \citep{arras2022variable, kim2024bayesian, kim2024imaging}.
Proximity to the information-theoretical optimum allows \resolve{} to generate superior imaging results compared to \clean{} characterized by higher resolution and dynamic range.
Another notable attribute lies in its ability to report uncertainties on the output images.
This is possible thanks to the usage of Metric Gaussian Variational Inference \citep{mgvi},
a computational scheme that is able to approximatively represent posterior probability distributions in very high dimensional settings.
The drawback of this more probabilistically rigorous approach lies in its extensive computational requirements, limiting its practical applicability to selected data sets.
Recently, the idea of a minor and major cycles as represented in modern versions of  \clean{} has been ported to \resolve{}, increasing its speed by orders of magnitude \citep{2024A&A...690A.387R}.

Regardless, there is an encouraging trend of increasing computational efficiency achieved through both hardware advancements and algorithmic insights in recent years.
Technical improvements originating from \resolve{} have been beneficial to the larger radio community, including those centered around \clean{}.
Key algorithmic advancements like the wgridder \citep{arras2021efficient} have been incorporated into various popular imaging packages, e.g., WSClean \citep{offringa2014wsclean}.

This paper extends the current capabilities of the \resolve{} algorithm to full polarization imaging.
The structure of this paper is as follows.
\Cref{sec:bayes_intro} summarizes the basics of Bayesian imaging.
In \cref{sec:derivation} we derive a generative model for polarized emission imaging.
Subsequently, we apply this model to a VLA Cygnus~A observation in \cref{sec:cygnusa}.
In \cref{sec:conclusion} we provide a summary of our findings.
In the interest of maintaining transparency throughout the entirety of this paper, any clipped color map is denoted by a color bar shown with a distinct triangular marker.

\section{Bayesian approach to polarization imaging}\label{sec:bayes_intro}
In contrast to \clean, the traditional imaging algorithm, our Bayesian approach solves radio interferometric imaging probabilistically.
While \clean{} computes just a single estimate of the sky, our algorithm infers the probability distribution of the Stokes~$I$, $Q$, $U$, and $V$ maps conditional to the measured data $d$.
Bayesian Stokes~$I$ imaging and calibration algorithms have already been developed in previous work building on the \resolve{} framework.
Recent examples are \citet{arras2021comparison} comparing \resolve{} with \clean{} and \citet{roth2023bayesian} introducing joint direction-dependent calibration and imaging.
In this article, we extend the \resolve{} framework to full Stokes imaging.
For completeness, we shortly recapitulate the general idea of Bayesian imaging.
A more thorough introduction can be found in previous work (e.g.\ \citealt{arras2021comparison}).

The probability distribution $P(I,Q,U,V|d)$ of the Stokes~$I$, $Q$, $U$, and $V$ maps conditioned on the measured data $d$ is called posterior distribution and can be expressed via Bayes' theorem:
\begin{align}
\mathcal P(I,Q,U,V|d) = \frac{\mathcal P(d|I,Q,U,V)\, \mathcal P (I,Q,U,V)}{\mathcal P (d|I,Q,U,V)}
\end{align}
in terms of the likelihood $\mathcal P (d|I,Q,U,V)$ and the prior $\mathcal P (I,Q,U,V)$.

The prior $\mathcal P(I,Q,U,V)$ encodes general knowledge by assigning a probability to each potential sky map.
Via this prior distribution, physical constraints can be encoded.
For example, in previous work on Stokes~$I$ imaging the prior distribution enforced the positivity of the sky brightness.
In \cref{sec:derivation}, we extend the Stokes~$I$ prior to full Stokes imaging.
The new prior model enforces consistency between the individual Stokes maps and especially ensures the fractional polarization to be between 0\% and 100\%.

The likelihood $\mathcal P (d|I,Q,U,V)$ specifies the probability of measuring the data $d$ for a given realization of the sky maps $I, Q, U$ and $V$.
For radio interferometry, it can often be assumed that the noise on the data $d$ is drawn from a Gaussian distribution with diagonal covariance.
Here, the term noise not only means the receiver noise, but also radio frequency interference imprints that have not been flagged, the net effect of gain inaccuracies, and the like.
The assumption of Gaussian statistics is then often justified approximately thanks to the independence of these contributions and the central limit theorem.

In practice, starting from a given set of Stokes sky maps, the corresponding model visibilities are computed via the radio interferometric measurement equation (RIME, \citealt{smirnov11}).
Then, the noise-weighted residual between the measured data and the model visibilities determines the normalized log-likelihood probability.

Applying Bayesian inference to real-world radio interferometric data comes with two challenges.
First, a physically meaningful but sufficiently flexible prior probability density for the Stokes sky maps will have a complicated form.
Second, since the sky maps typically have millions of pixels, the algorithm used for approximating the posterior distribution needs to be numerically efficient and capable of dealing with high dimensional parameter spaces.

To address the first challenge, we do not directly set up a prior distribution in the sky map domain but instead encode the prior in the form of a normalized generative model \citep{Knollmueller2018}.
This means we derive a physics-inspired model mapping independently Gaussian distributed random numbers $(\xi)$, called latent parameters, onto the desired correlated distributions of the Stokes~$I$, $Q$, $U$, and $V$ maps.
This generative model encoding the prior distribution is derived in \cref{sec:derivation}.

To deal with the second challenge, the high dimensional parameter space, we rely on variational inference to find an approximation of the posterior distribution.
More specifically, we employ the MGVI algorithm \citep{mgvi} to infer the posterior distribution.
In essence, the MGVI algorithm finds a Gaussian approximation to the posterior distribution of the latent parameters $\xi$, allowing not only to access the posterior mean sky maps but also their uncertainties.
For the details of the MGVI algorithm, see \citet{mgvi} and \citet{arras2021comparison}, where it was already applied to radio interferometric imaging.

\section{Derivation of model}\label{sec:derivation}
Generally speaking, the polarization state of a monochromatic electromagnetic wave can be expressed with the help of the four \emph{Stokes parameters} \citep{stokes1851}: $I$, $Q$, $U$ and $V$.
They denote the absolute intensity, the two linear polarization degrees of freedom, and the circular polarization, respectively.

The polarized sky brightness distribution is a complex $2\times2$ matrix:
\begin{align}
\label{eq:intro:X}
X =\begin{pmatrix} \langle e_{a,l}e_{b,l}^{*}\rangle &\langle e_{a,l}e_{b,r}^{*}\rangle\\\langle e_{a,r}e_{b,l}^{*}\rangle &\langle e_{a,r}r_{b,r}^{*}\rangle \end{pmatrix} = \begin{pmatrix} I-V & Q+\iu U\\Q-\iu U & I+V \end{pmatrix}
\end{align}
in circular basis, that is the electromagnetic field $e$ is measured with circular feeds, and
\begin{align}
X =\begin{pmatrix} \langle e_{a,x}e_{b,x}^{*}\rangle &\langle e_{a,x}e_{b,y}^{*}\rangle\\\langle e_{a,y}e_{b,x}^{*}\rangle &\langle e_{a,y}r_{b,y}^{*}\rangle \end{pmatrix}= \begin{pmatrix} I+Q & U+\iu V\\U-\iu V & I-Q \end{pmatrix}
\end{align}
in linear basis \citep{smirnov11}.
We follow here the physics convention to specify circular polarization and note that the astronomical convention differs from this in the sign of $V$.
The indices $a,b$ are antenna labels and the indices $l,r$ and $x,y$ refer to the circular and linear feeds, respectively.

Crucially, the matrix $X$ has to satisfy three constraints in order to be physically sensible:
\begin{enumerate}
\item\label{it:intro:cond1} $X$ is positive definite and Hermitian.
\item\label{it:intro:cond3} The total brightness $I$ is strictly positive: $I>0$.
\item\label{it:intro:cond2} The polarized part of the emission cannot exceed the Stokes~$I$ emission:
  \begin{align}\label{eq:intro:ineq2}
  I\geq\sqrt{Q^{2}+U^{2}+V^{2}}.
  \end{align}
\end{enumerate}
More details on Stokes parameters can be found in, e.g., \citet{hamaker96,smirnov11}.

In our previous work \citep{arras2021comparison}, the basic idea for Stokes-$I$ imaging was to model the sky brightness distribution $I$ with an exponentiated Gaussian process $s$:
\begin{align}\label{eq:intro:exp}
I = e^{s}
\end{align}
Now, we generalize this approach to polarization imaging.
The goal is to write down a generative model that is a function of standard normal distributed parameters $\xi$ to Stokes~$I$, $Q$, $U$ and $V$ such that it is guaranteed that the above criteria are met.

The crucial idea for the polarization model is to generalize \cref{eq:intro:exp} to matrix form and express the $X$ as matrix exponential:
\begin{align}\label{eq:intro:polform}
X =e^{x} \coloneqq \exp \begin{pmatrix} s+q & u+\iu v\\u-\iu v & s-q \end{pmatrix},
\end{align}
where $s, q, u$, and $v$ are real numbers for each pixel, in particular they can be both positive and negative.

Let us verify that \cref{eq:intro:polform} indeed satisfies the above conditions.
From the fact that Hermitian conjugation and exponentiation of a matrix commute and since $x$ is Hermitian, $e^{x}$ is Hermitian as well and thereby has only real eigenvalues.
Since the eigenvalues of the exponential of a matrix are given by the exponentiated eigenvalues of the matrix and because $x$ has only real eigenvalues, $e^{x}$ is positive definite.
This proves condition~\ref{it:intro:cond1}.
For showing condition~\ref{it:intro:cond3}, \cref{eq:intro:X} and \cref{eq:intro:polform} need be combined to express $I, Q, U$ and $V$ in terms of $s, q, u$ and $v$:
\begin{align}
I & = e^{s} \cosh p,            & Q & = \frac{q}{p} e^{s}\sinh p, \label{eq:intro:pola1} \\
U & = \frac{u}{p} e^{s}\sinh p, & V & = \frac{v}{p} e^{s}\sinh p,\label{eq:intro:pola2}
\end{align}
with $p \coloneqq \sqrt{q^{2} + u^{2}+ v^{2}}$.
It is apparent that $I>0$ is naturally guaranteed in this formulation.
Condition~\ref{it:intro:cond2} (\cref{eq:intro:ineq2}) holds as well because $e^{x}$ has only positive eigenvalues.
Therefore, the determinant, which is the product of the eigenvalues, is positive:
\begin{align}\label{eq:intro:poladet}
0 < \det X = I^{2} - Q^{2}-U^{2}-V^{2}.
\end{align}
Since $I>0$, there is no sign ambiguity and \cref{eq:intro:poladet} is indeed equivalent to condition~\ref{it:intro:cond2}.
Thus, all three conditions are fulfilled.

All in all, this approach provides a natural way to model polarized emission of, for instance, radio sources.
Its major advantages are that it correlates the Stokes~$I$ and the Stokes~$Q$, $U$, and $V$ components in a non-trivial yet natural way.
Additionally, it ensures that the polarized emission cannot exceed the Stokes~$I$ component and that the Stokes~$I$ component is strictly positive.
Both are physical constraints that are strictly speaking necessary to build into an imaging algorithm because as soon as these constraints are violated the result of the imaging algorithm is definitely not a faithful representation of physical reality.


\section{Application to radio interferometric data}\label{sec:cygnusa}

\subsection{Data model and prior probability density}

To show that the above presented prior model provides additional value in actual applications, we demonstrate how it performs on VLA data.
We choose the same data set that was used in \citet{sebokolodi2020wideband} and \citet{arras2021comparison} to be able to compare traditional \clean{} imaging with a \resolve{} reconstruction based on our polarization prior model.
Additionally, we can compare the new \resolve{} Stokes~$I$ reconstruction to the previous ones.

All in all, the setup of the reconstruction is very similar to the previous Stokes~$I$ \resolve{} reconstruction by \citet{arras2021comparison}.
The main difference is that we now consider full polarization imaging, which turns every pixel of the sky into a complex-valued $2\times2$ matrix $X$ defined in \cref{eq:intro:X}, which the Stokes parameters determine the the real and imaginary parts.
As in the earlier work, the sky brightness has two components, one modeling diffuse emission and the other point sources.
In this work, we have extended both components to full polarization using the model outlined in \cref{sec:derivation}.

More specifically, in the previous diffuse sky model, we had a single Gaussian process $s$ modeling the Stokes-$I$ diffuse emission (see \cref{eq:intro:exp}).
Now we have four Gaussian processes $s, q, u$ and $v$ modeling all Stokes components of the diffuse emission as defined in \cref{eq:intro:polform}.
As in \citet{arras2021comparison}, the Gaussian processes are encoded in the form of generative models mapping standard normal distributed latent parameters $\xi_s,\, \xi_q,\, \xi_u,\, \xi_v$ to the actual Gaussian process values $s(\xi_s),\, q(\xi_q),\, u(\xi_u),\, v(\xi_v)$.
The exact mathematical definition of the generative Gaussian process model can be found in \citet[sec.~3.4]{arras2021comparison}.
In \cref{tab:hpres} we report the chosen hyperparameters for the Gaussian processes.

In alignment with the previous Stokes~$I$ imaging paper, we employ an inverse gamma model (IGM) for Stokes~$I$ to model the two known point sources within the field.
This is further supplemented with consistent Stokes~$Q$, $U$, and $V$ components.
The point sources are strategically positioned at the phase center---corresponding to the primary AGN---and at coordinates $(\ang{;;0.35},\ang{;;-0.22})$ associated with a second point source, called the secondary transient \citep{perleytransient}.

For the prior parameters, we employ $\alpha_\text{IGM}=0.5, q_\text{IGM}=0.2$ for the inverse-gamma Stokes~$I$ prior, while utilizing zero and one as the mean and standard deviation, respectively, for the $q, u$ and $v$ parameters.
As the diffuse emission model, the point source model is coded as a generative model mapping standard normal distributed latent parameters $\xi^\text{p}$ to the actual full Stokes point source sky brightness $X^\text{p}(\xi^\text{p})$.

To summarize, our sky model has two components, one for diffuse emission $X^\text{d}$ and the other for point sources $X^\text{p}$.
Each component described by a generative model mapping latent parameters $\xi^\text{d/p}=(\xi_s^\text{d/p},\, \xi_q^\text{d/p},\, \xi_u^\text{d/p},\, \xi_v^\text{d/p})$ to the full Stokes sky brightness.
The total sky brightness
\begin{align}
X(\xi^\text{d}, \xi^\text{p}) = X^\text{d}(\xi^\text{d}) + X^\text{p}(\xi^\text{p})
\end{align}
is the sum of the two components.
The full Stokes sky brightness is a complex-valued $2 \times 2$ matrix at every sky location.
Thus, when we discretize the sky with $M \times M$ pixels, the sky brightness $X \in {\left(\mathbb{C}^2 \times \mathbb{C}^2\right)}^{M \times M}$ is a complex-valued $2 \times 2$ matrix for each of the sky pixels.

For full Stokes imaging the data points are also complex-valued $2\times2$ matrices. Thus mathematically each data point $d^{i}$, commonly named visibility, has the form
\begin{align}
    d^{i} = \begin{pmatrix} d_{ll}^{i} & d_{lr}^{i}\\ d_{rl}^{i} & d_{rr}^{i} \end{pmatrix}.
\end{align}
With $d = \left(d^{0}, ..., d^{L}\right)$, we denote the vector containing all visibilities of the observation.
In this notation, the radio interferometric measurement equation \citep{smirnov11} turns into:
\begin{align}
    d = R\left(X\right) + n,
\end{align}
with $R:{\left( \mathbb{C}^2 \times \mathbb{C}^2 \right)}^{M \times M} \to {\left( \mathbb{C}^2 \times\mathbb{C}^2 \right)}^L$ being the measurement operation, and $n \in {\left( \mathbb{C}^2 \times\mathbb{C}^2 \right)}^L$ the measurement noise.
For the noise, we assume Gaussian statistics. Nevertheless, we do not assume a fixed noise covariance but reconstruct it simultaneously with the sky brightness.
Also, the prior for the noise covariance is encoded as a generative model mapping $\xi^\text{n}$ to the noise covariance $N(\xi^\text{n})$.
Thus, our likelihood takes the form:
\begin{align}
\mathcal P (d | \xi^\text{s}, \xi^\text{n}) = \mathscr G \left(d - R \left(X(\xi^\text{s})\right), N(\xi^\text{n})\right),
\end{align}
with $X(\xi^\text{s})$ being the generative model for the sky brightness (with $\xi^\text{s} = (\xi^\text{d}, \xi^\text{p})$), and $N(\xi^\text{n})$ the generative model for the noise covariance.
For the noise covariance, we use the same model already presented in \citet{arras2021comparison}.
In essence, it uses the weights of visibilities recorded with the data as the inverse noise covariance and corrects them by an additional factor depending on the baseline length.
This baseline-length-dependent additional noise factor is reconstructed simultaneously with the sky brightness.
In this specific case, the reported weights were all ones, thus the inferred noise levels are completely reflecting the degree to which the sky model is able to fit the data as a function of baseline.
More details about the noise model can be found in its original description in \citet{arras2021comparison}.

\begin{table*}
\centering
\begin{tabular}{lrrrrrr}
\toprule
& $\alpha$ mean & $\alpha$ sd &$s$ mean & $s$ sd & $q, u, v$ mean & $q, u, v$ sd\\\midrule
Offset&0&---& 21 & --- & 0 & ---\\
{[1]} Zero mode variance & 2 & 2 & 1 & 0.1 & 0.01 & 0.01 \\
{[2]} Fluctuations & 2 & 2 & 5 & 1 & 0.01 & 0.01 \\
{[5]} Flexibility & 1.2 & 0.4 & 1.2 & 0.4 & 0.1 & 0.1 \\
{[6]} Asperity & 0.2 & 0.2 & 0.2 & 0.2 & 0.2 & 0.2 \\
{[7]} Average slope & -2 & 0.5 & -2 & 0.5 & -2 & 0.5 \\\bottomrule
\end{tabular}
\caption{Hyper parameters for \resolve{} runs. Analogous to \citet[Tab.\ 1]{arras2021comparison}.}\label{tab:hpres}
\end{table*}

We publish all \clean{} maps and all \resolve{} approximate posterior samples for the four frequencies 2052~MHz, 4811~MHz, 8427~MHz and 13360~MHz on zenodo \citep{zenodo} as hdf5 and FITS files.

\subsection{Outline of \resolve{} reconstruction and initial comparison to \clean}

\begin{figure*}
\centering
\input{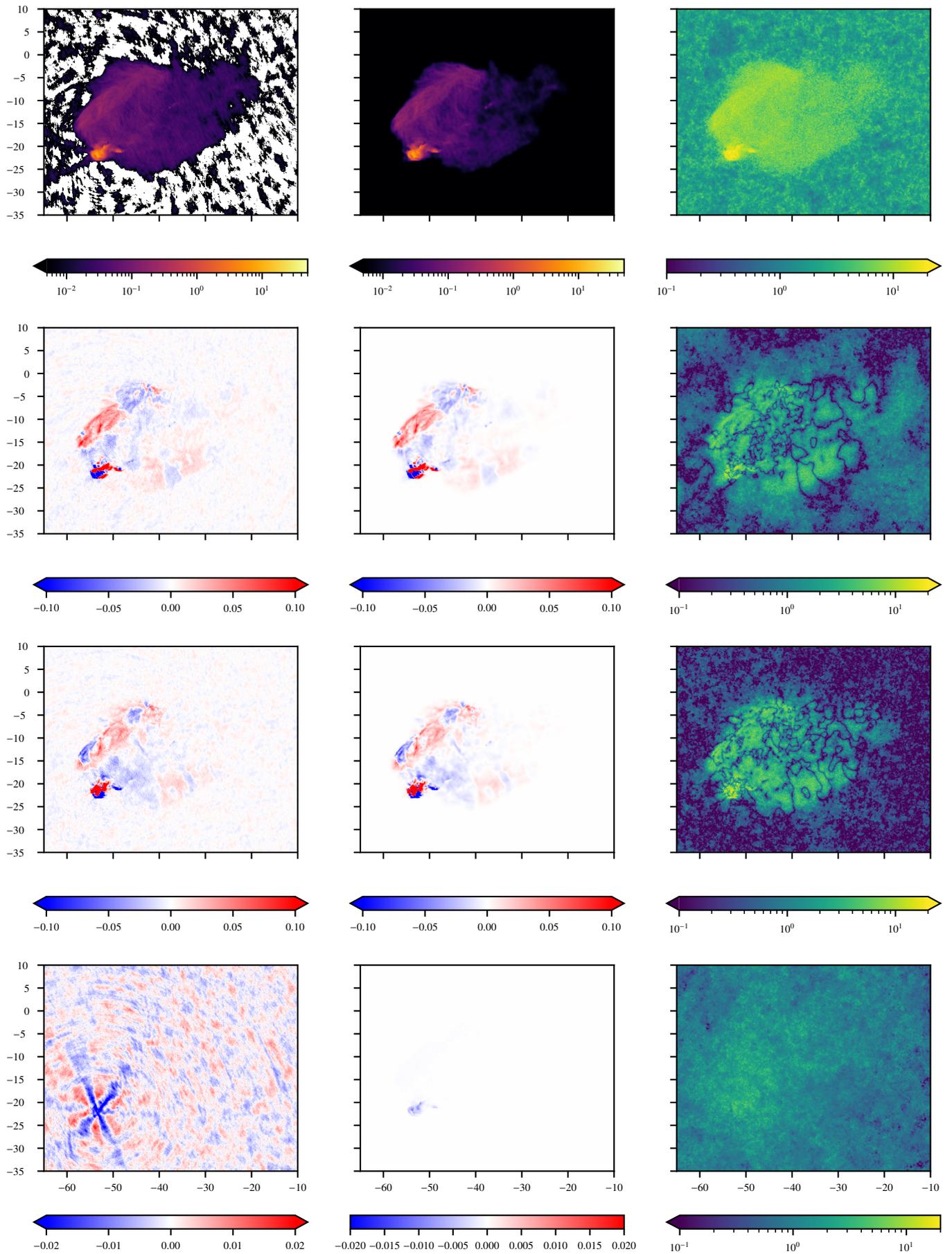}
\caption{
Overview of the Eastern lobe in \SI{13}{\giga\hertz}.
From top to bottom, the rows respectively represent Stokes~$I$, $Q$, $U$, and $V$ emissions in \unit{\jy\per\as\squared}.
Spanning left to right, the columns illustrate the \clean{} restored image, the \resolve{} posterior mean $m_1$, and the quotient $|m_1| / \sigma_1$, where $\sigma_1$ is the \resolve{} posterior standard deviation.
}\label{fig:lefthotspot}
\end{figure*}

In order to demonstrate the advantages and characteristics of our proposed algorithm, we apply it to VLA Cygnus~A observations.
We choose the data sets that \citet{sebokolodi2020wideband} and \citet{arras2021comparison} image as well.
These data sets contain four channels with frequencies of approximately \SIlist{2;4;8;13}{\giga\hertz}.
We focus the discussion on the analysis of the \SIlist{2;13}{\giga\hertz} reconstructions, as these represent the most extreme scenarios.
The full images including posterior samples are provided as supplementary material \citep{zenodo}.

\Cref{fig:lefthotspot} juxtaposes \clean{} and \resolve{} reconstructions of the same \SI{13}{\giga\hertz} VLA dataset focusing on the Eastern Lobe.
As visible in the first row the \clean{} and \resolve{} Stokes~$I$ reconstructions agree well in regions with significant surface brightness.
In contrast, in low surface brightness regions, the \clean{} maps suggest negative-brightness zones, which are in contradiction to physical laws.
\resolve{} does not generate these unphysical negative-brightness features.

The rightmost column shows a map of the relative certainty, which is the posterior mean normalized by the posterior uncertainty on a pixel-by-pixel basis.
High-brightness regions have values exceeding 20 on the relative certainty map, indicating uncertainties of less than 5\%.
Conversely, in the minimal to no-brightness regions, the relative certainty falls below one, reflecting uncertainties exceeding 100\%.
Importantly, these proportions align with the constraint that ensures Stokes~$I$ remains positive.

Subsequently, the second and third rows present Stokes~$Q$ and $U$, respectively, where the \clean{} and \resolve{} reconstructions largely coincide.
Two key differences, however, are noteworthy.
Firstly, \clean{} unveils significant linearly polarized emissions in regions devoid of substantial Stokes~$I$ brightness---a circumstance that is unphysical.
It can be substantiated that the polarization consistency constraints are compromised in these regions.
Secondly, the \resolve{} reconstructions appears smoother, contain fewer artifacts, and provide more insights into both low-brightness and high-brightness regions.
The relative certainty maps for Stokes~$Q$ and $U$ show lines of zero values caused by the respective quantity switching sign at these locations.

Finally, we turn our attention to the Stokes~$V$ reconstructions.
Notably, the \clean{} algorithm exhibits significant positive and negative Stokes~V emission in a radial pattern that emanates from the hotspot; these are clearly artefacts.
Contrarily, the \resolve{} method presents negligible negative Stokes~V emission centered on the hotspot (note the different colour bar) and virtually no emission in the remaining regions.
Importantly, the relative certainty map underscores that this minimal emission is not significant.
We therefore decide to exclude Stokes~V from the subsequent analysis in this article.

\begin{figure*}
\centering
\input{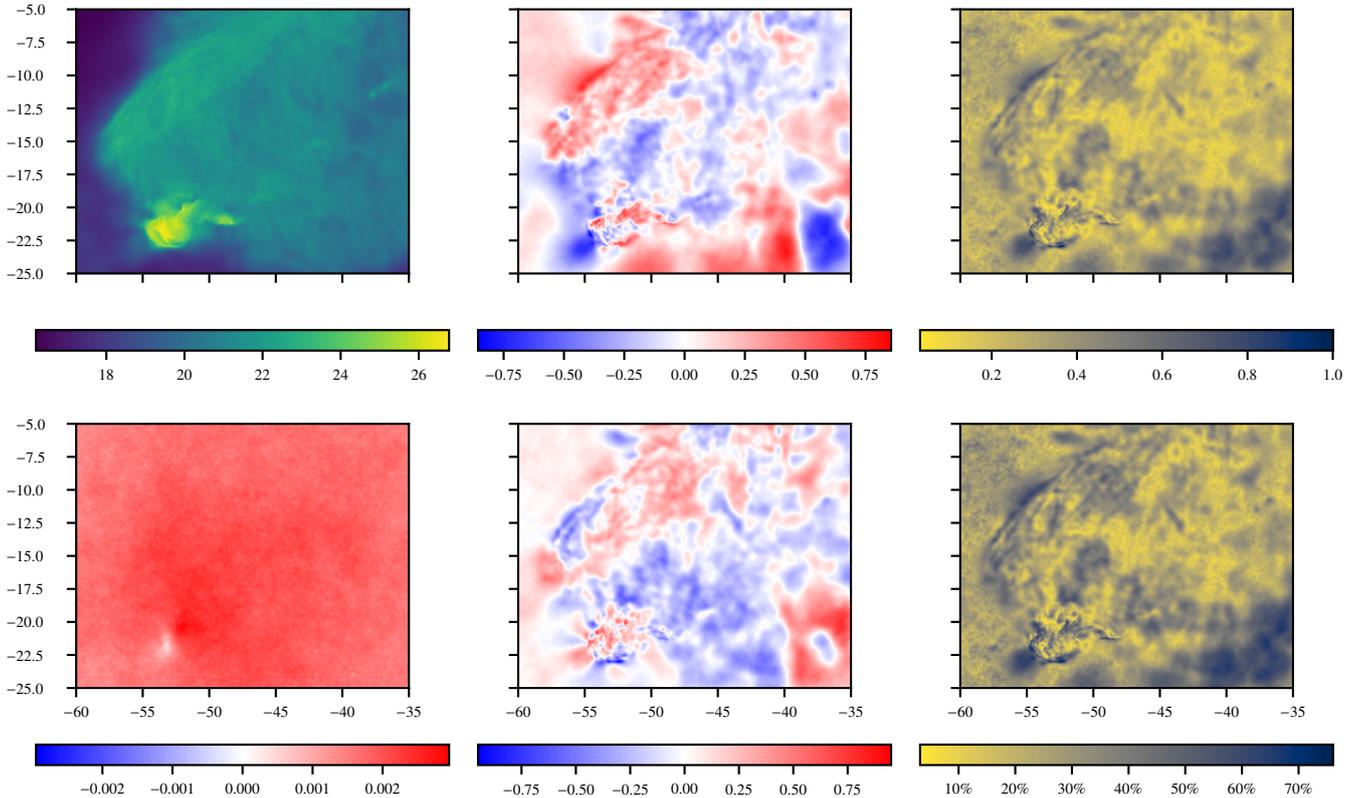}
\caption{From left to right and top to bottom: \resolve{} posterior mean of the unitless parameters $s$, $q$, $p$, $v$, $u$ and $\tanh p$ of the \SI{13}{\giga\hertz} reconstruction.}
\label{fig:logquantities}
\end{figure*}

For reference and to build some intuition, we display the posterior means of the parameters $s$, $q$, $u$ and $v$ and the derived quantity $p$ in \cref{fig:logquantities}.

\begin{table}
\begin{tabular}{lcc}
\toprule
Freq [MHz]& $\chi^2_{\resolve{}}$ & $\chi^2_{\clean{}}$\\
\midrule
2052  & $1.02 \pm 0.24$ & $115.32$ \\
4811  & $1.06 \pm 0.15$ & $38.83$ \\
8427  & $1.06 \pm 0.02$ & $51.35$ \\
13360 & $1.00 \pm 0.02$ & $38.62$ \\
\bottomrule
\end{tabular}

\caption{
Reduced $\chi^2$ values of the \resolve{} and \clean{} reconstructions of the four frequencies.
As \resolve{} provides posterior samples the $\chi^2$ mean and standard deviation of the samples is indicated.
}\label{tab:redchisq}
\end{table}

\begin{figure*}
\centering
\input{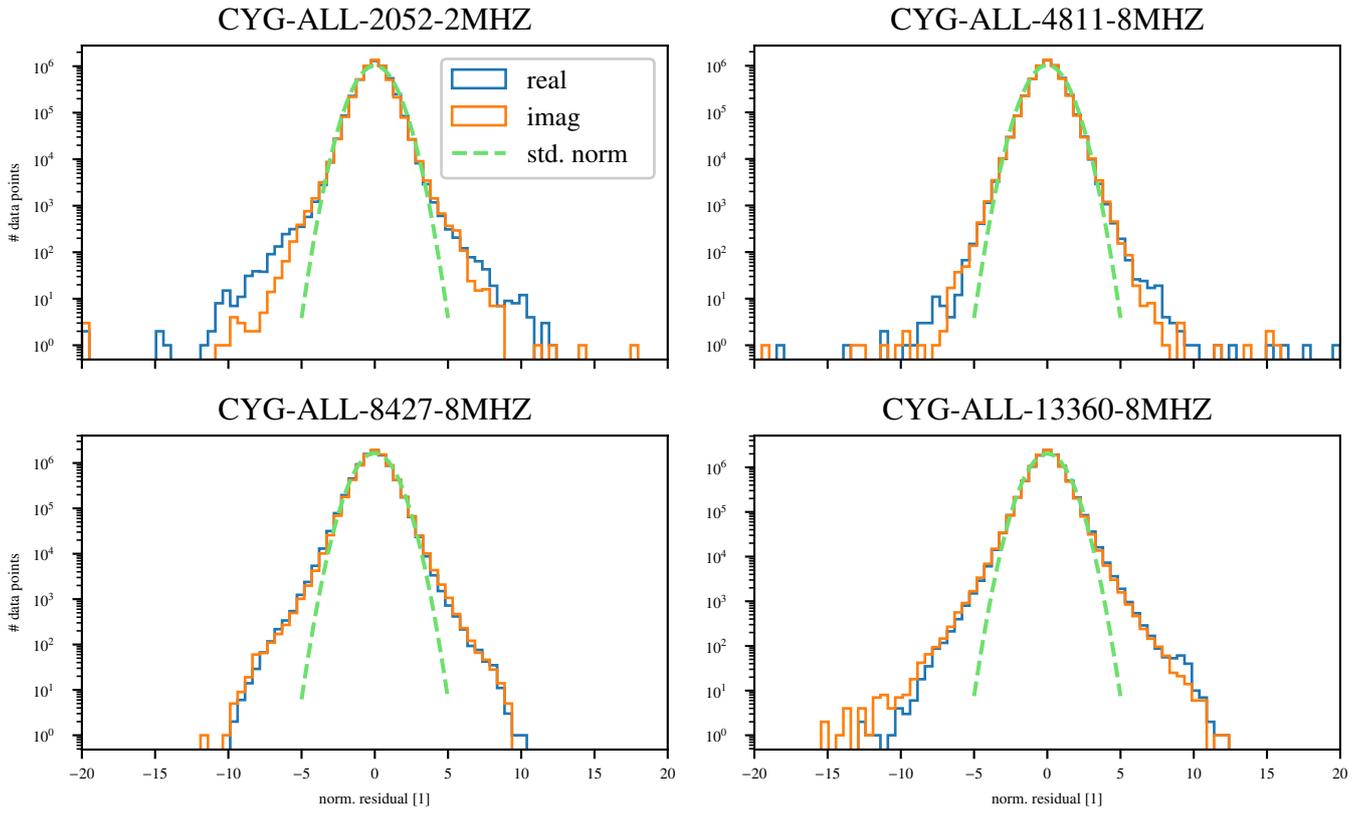}
\caption{Normalized residuals of all posterior samples.}
\label{fig:normresiduals}
\end{figure*}

\begin{figure*}
\centering
\loadplotpng{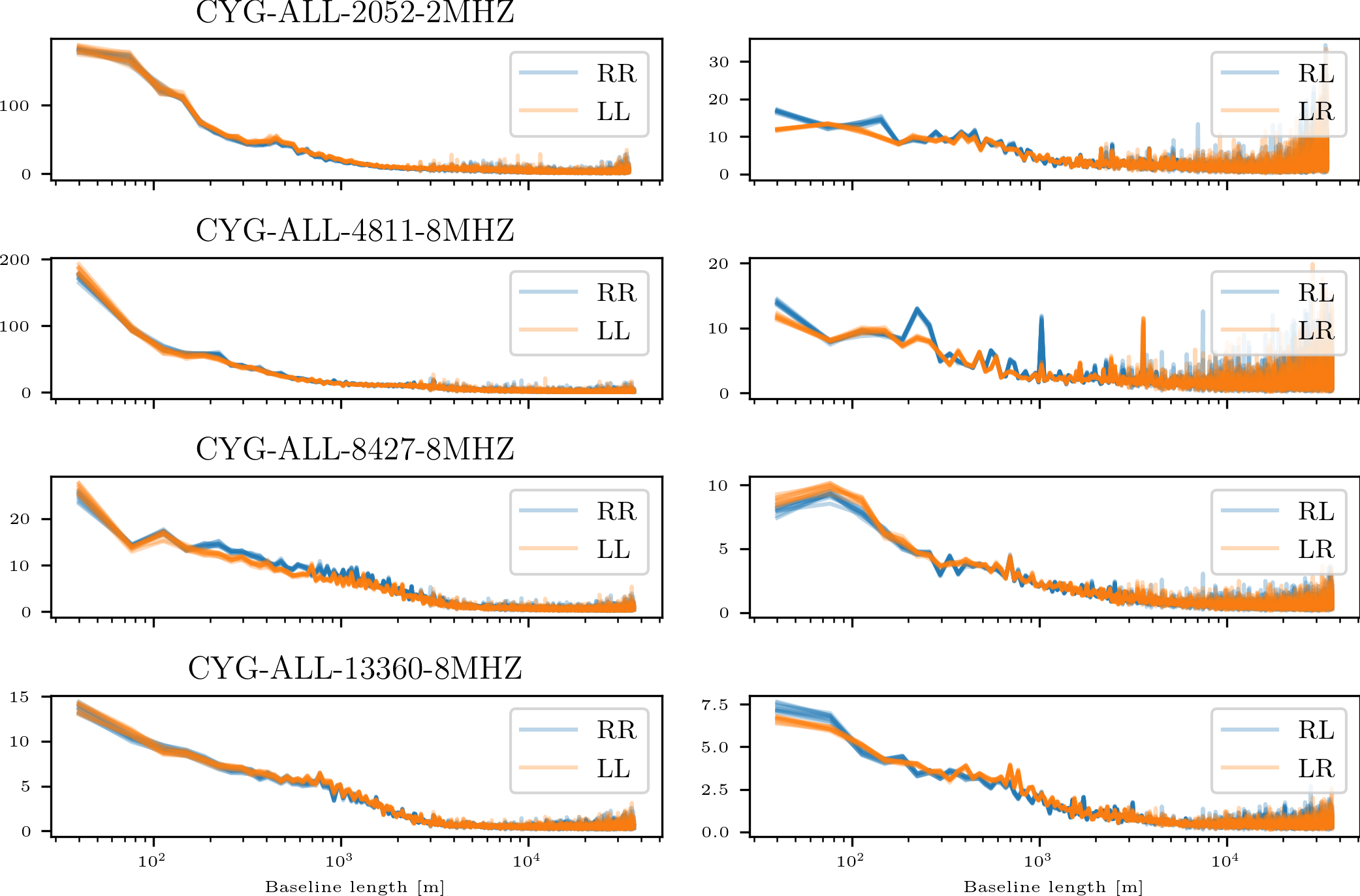}
\caption{Posterior samples of the correction function as part of the Bayesian weighting scheme $N(\xi^\text{n})$. This quantity is unitless.}
\label{fig:sigmacorr}
\end{figure*}

We assess the reconstruction quality by checking compatibility with the data.
Inconsistencies between data and reconstruction indicate possible problems such as unmodeled systematic effects, inaccurate prior modeling of the sky brightness distribution, or an unconverged numerical algorithm, among other potential factors.
Data compatibility can be quantified with the reduced $\chi^{2}$ value, defined as
\begin{align}
  \chi^{2}_{\mathrm{red}}= \frac{|\sqrt{N^{-1}}  (d-Rs)|^2}{n},
\end{align}
where here $n$ represents the number of real degrees of freedom in the data (and not the noise).
Ideally, if the reconstruction matches perfectly with the data $d$ and associated reported noise levels $N$, $\chi^{2}_{\mathrm{red}}$ equals 1.
The reduced $\chi^{2}$ values of the \resolve{} and \clean{} reconstructions are displayed in \cref{tab:redchisq}.
For both algorithms the $\chi^{2}$ values where computed using the noise covariance inferred by \resolve{}.
Given that we are not working with a single value, but a collection of posterior samples, both the posterior mean and standard deviation are presented for a comprehensive assessment.
The posterior means are close to 1, as predicted, and the posterior standard deviation is comparably small (see \cref{fig:normresiduals}).
This outcome may not be surprising, given that we reconstruct the noise level in tandem with the images.
For a detailed discourse on why fitting the noise level is necessary, we refer to \citet{arras2021comparison}.
The results of our noise fitting scheme (also known as Bayesian weighting scheme) are displayed in \cref{fig:sigmacorr}.
Similar to \citet{arras2021comparison}, we observe that short baselines contain significantly more systematic errors than advertised by the noise level reported in the data. Therefore the correction function increases for small $|k|$.
This effect is the stronger the lower the observing frequency is.
In contrast to the \resolve{} posterior samples, the \clean{} reconstructions have reduced $\chi^{2}$ values significantly larger than one, indicating that the \clean{} images are less compatible with the data.

In summary, the comparison reveals superior image quality and data fidelity of the \resolve{} reconstructions compared to the \clean{} maps.

\begin{figure*}
\centering
\input{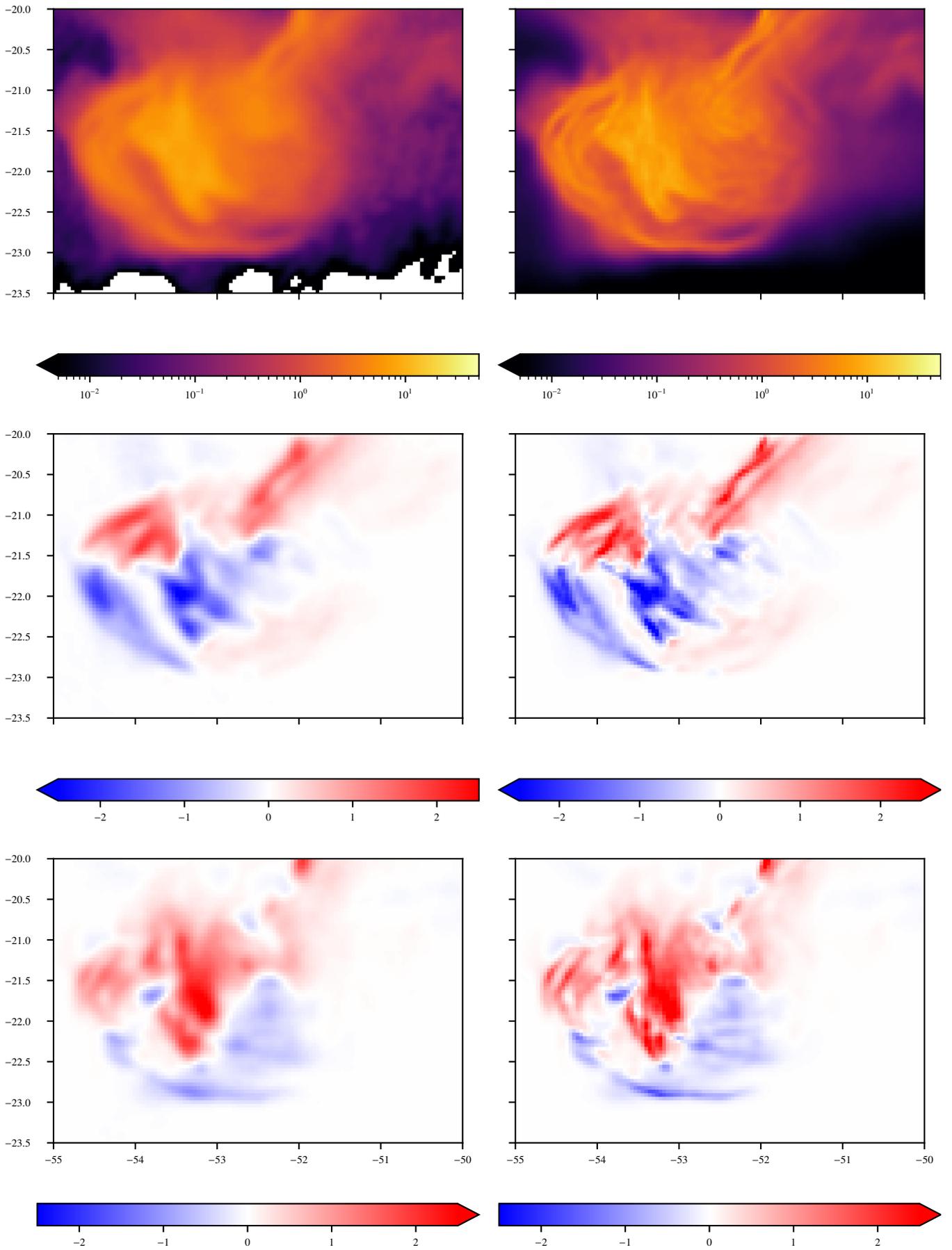}
\caption{
Amplified view of the Eastern hotspot in \SI{13}{\giga\hertz}, exemplifying a region with high signal-to-noise.
From top to bottom, the rows respectively represent Stokes~$I$, $Q$, and $U$ emissions in \unit{\jy\per\as\squared}.
Spanning left to right, the columns illustrate the \clean{} restored image and the \resolve{} posterior mean.
}\label{fig:lefthotspot_highsnr}
\end{figure*}

For further analysis, we investigate high brightness regions, consequently having high signal-to-noise ratios.
For this we focus on the Eastern hotspot of Cygnus~A.
\Cref{fig:lefthotspot_highsnr} juxtaposes \clean{} and \resolve{} Stokes~$I$, $Q$, and $U$ reconstructions.
Again they agree well with each other, with the noteworthy distinction that \resolve{} exhibits a significantly higher resolution than \clean{}.
This closely mirrors the findings reported by \citet{arras2021comparison} wherein higher resolution images were obtained with \resolve{}.
One explanation for this stems from its ability to bypass a convolution with the restoring beam that is here applied to the \clean{} component image according to the traditional way \clean{} results are presented.
A map of not convolved \clean{} components would, however, exhibit almost zero flux in between those, even in regions of diffuse emission, which is not very physical.
With \citet{arras2021comparison} exclusively considering Stokes~$I$, here we extend this observation to full polarization imaging.

As stated earlier, in terms of $\chi^{2}$, the \resolve{} reconstructions are more compatible with the data than \clean{} ones.
This suggests that these structures indeed mirror reality more closely.
A confirmation that many of the structures revealed by \resolve{} are real was given by \citet{arras2021comparison} for Stokes $I$ by showing that higher frequency and therefore higher resolution \clean{} maps confirm the structures seen by \resolve{}.

In summary, in regions with high signal-to-noise ratios, \resolve{} has the potential to extract higher resolution images from the same data set.

\begin{figure*}
\centering
\input{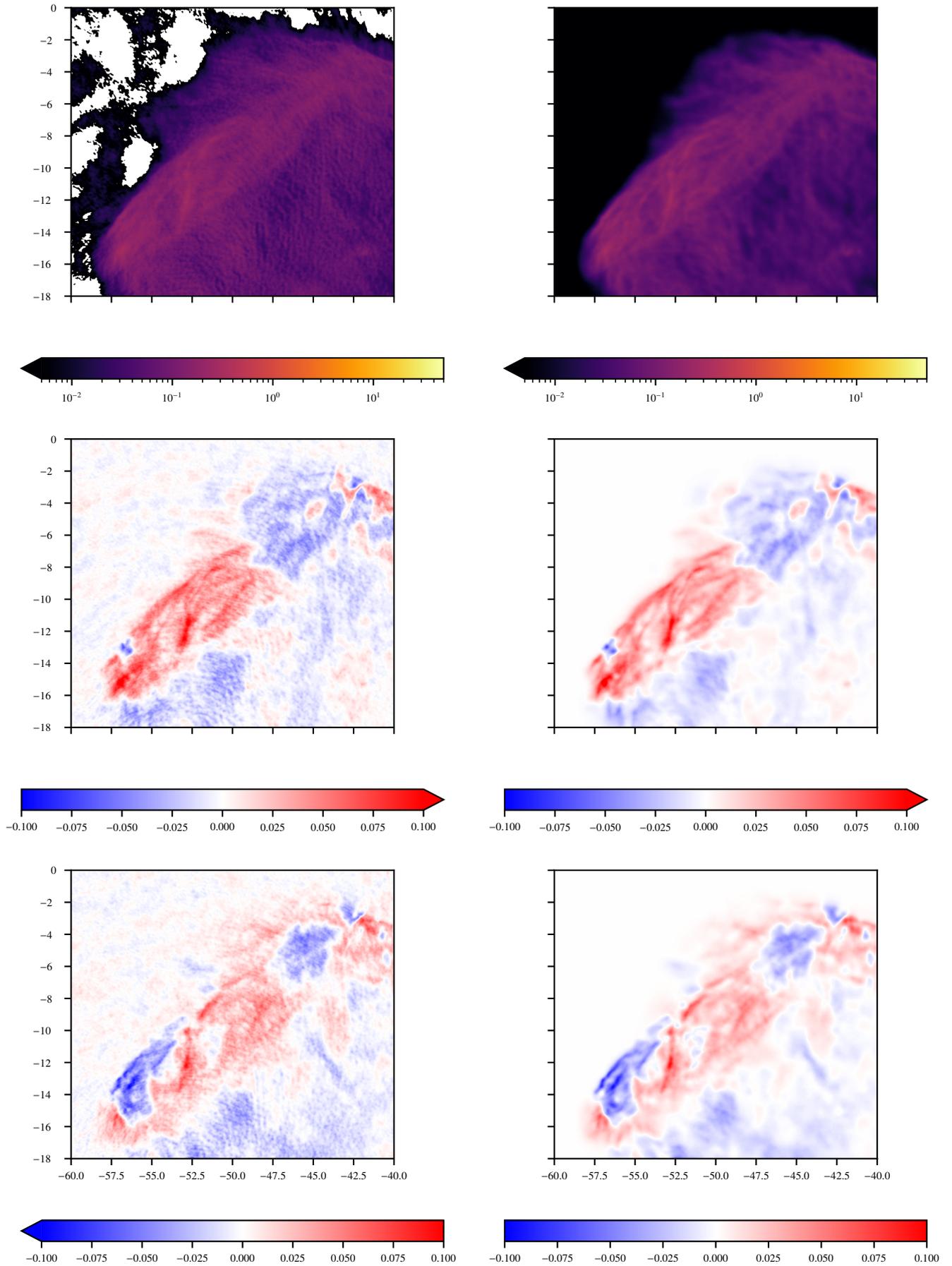}
\caption{
Amplified view of the Northern flank of the Eastern lobe, exemplifying a region characterized by a low signal-to-noise ratio.
From top to bottom, the rows respectively represent Stokes~$I$, $Q$, and $U$ emissions \unit{\jy\per\as\squared}.
Spanning left to right, the columns illustrate the \clean{} restored image, the \resolve{} posterior mean $m_1$, and the quotient $|m_1| / \sigma_1$, where $\sigma_1$ corresponds to the \resolve{} standard deviation.}\label{fig:lefthotspot_lowsnr}
\end{figure*}

Next, we analyze a prototypical region characterized by a low signal-to-noise ratio, specifically the Northern portion of the Eastern lobe, as depicted in \cref{fig:lefthotspot_lowsnr}.
The first row presents the Stokes~$I$ emissions for both \clean{} (left column) and \resolve{} (right column).
One obvious observation is the identification of white areas in the \clean{} image that lie in close proximity to regions of emission.
These are unphysical negative brightness regions.
As previously discussed, \resolve{} excludes negative Stokes~$I$ brightness a priori, which eliminates the occurrence of white spots in the corresponding \resolve{} image.

Upon closer inspection of the emission regions, the \clean{} image shows a small-scale structure featuring a distinct spatial frequency uniformly across the image.
In stark contrast, the \resolve{} image is significantly smoother, with no distinguishing spatial frequencies.

Finally, we consider the linear polarization maps.
Again, the \clean{} and \resolve{} reconstructions are coherent in regions of significant emission, with \resolve{} having superior resolution in high-brightness regions and a considerably smoother texture in low-brightness regions.
Notably, the \clean{} image displays linear polarization in regions where negative brightness exists in the Stokes~$I$ image.
This is clearly unphysical.
In contrast, the \resolve{} reconstruction appropriately shows a lack of visible linear polarisation in regions of minimal flux, a result warranted by the prior.

In conclusion, we observe that \resolve{} exhibits enhanced performance in both low and high signal-to-noise regions.
The \resolve{} images are smoother in low signal-to-noise regions, while the images have superior resolution in high signal-to-noise environments.
\clean{} is in principle capable of producing images with varying resolution, either via showing the \clean{} component map or via its multi-scale version.
The former, however, shows unphysical structures for diffuse emission fields, vanishing flux areas between the \clean{} components.
This is why basically all presentations of its results, as the one in this publication, are convolved with the restoring beam as the final imaging step.
The latter, multi-scale \clean{}, provides varying resolution, but does not seem to reach the sub-beam resolution that \resolve{} exhibits, at least not in the case studied in \cite{arras2021comparison}.

\begin{figure}
\centering
\input{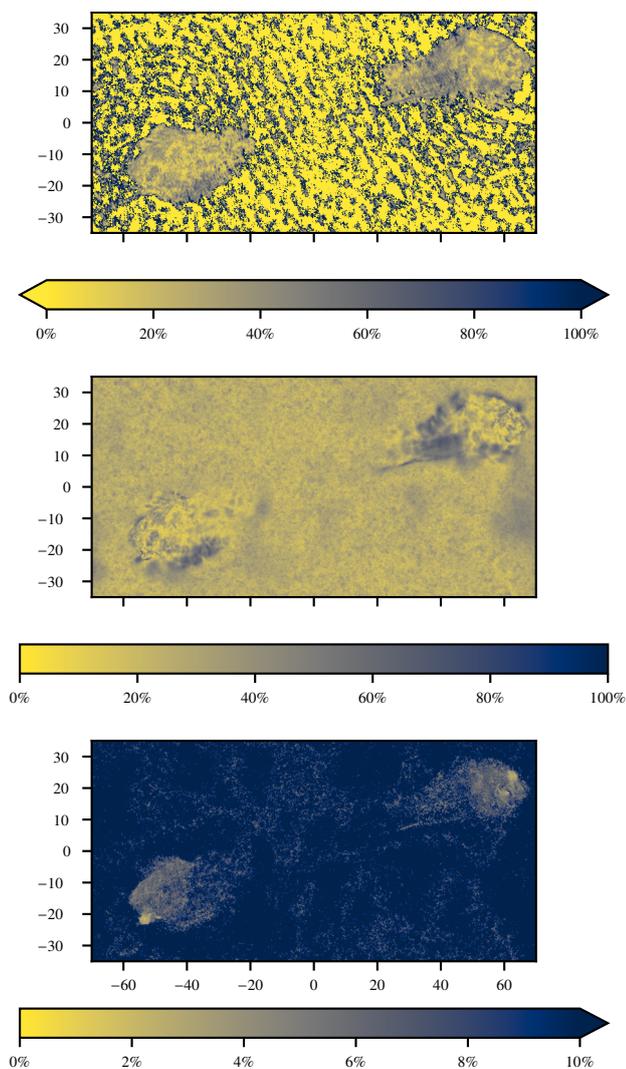}
\caption{
Fractional linear polarization.
From top to bottom: \clean\ map, \resolve{} posterior mean, \resolve{} posterior standard deviation.
}\label{fig:linpol}
\end{figure}

As a next step, we discuss fractional linear polarization, a parameter often used in scientific investigations, defined as $\sqrt{Q^{2} + U^{2}} / I$.
In \cref{fig:linpol}, the first row depicts the fractional linear polarization generated by \clean{}.\@
It is crucial to note that the color bar had to be adjusted due to the presence of values both below 0\% and above 100\% for linear polarization.
This unphysical result is fundamentally the consequence of \clean{} generating the images independently from each other, thereby failing to enforce the polarization consistency constraints.

The second row of \cref{fig:linpol} illustrates the fractional linear polarization computed from the \resolve{} outputs.
Importantly, the fractional linear polarization depends non-linearly on the Stokes parameters.
Therefore, the fractional linear polarization must be computed for each sample before averaging.
Expressed in formulas, we compute the mean and similarly the standard deviation, via  $\left\langle\sqrt{Q^{2} + U^{2}} / I\right\rangle_{(I,Q,U,V|d)}$, with $\langle f(x) \rangle_{(x|d)} := \int\mathcal D x \,\mathcal P(x|d)\, f(x)$ denoting the expectation value of the observable $f(x)$ with respect to the Bayesian posterior $\mathcal P(x|d)$ for $x$.

Evidently, the fractional polarization computed from the \resolve{} results maintains well within the limits of 0\% and 100\% as anticipated.
A closer inspection of the last row in \cref{fig:linpol} reveals that regions devoid of significant brightness exhibit correspondingly high standard deviation values, signaling maximal uncertainty in the determinant fractional linear polarization there.
Conversely, in the high brightness regions, the uncertainty in the fractional linear polarisation is less than 1\%, indicating a high degree of confidence of the algorithm.

Undoubtedly, upon comparing \clean{} and \resolve\, a cursory examination suggests that their reconstructions seem to align.
However, this compatibility is less apparent when comparing more closely, especially considering that \resolve{} does not exhibit the regions with negative fractional polarization encountered in the \clean{} reconstruction.

\subsection{Uncertainty depolarization}

\begin{figure}
\centering
\input{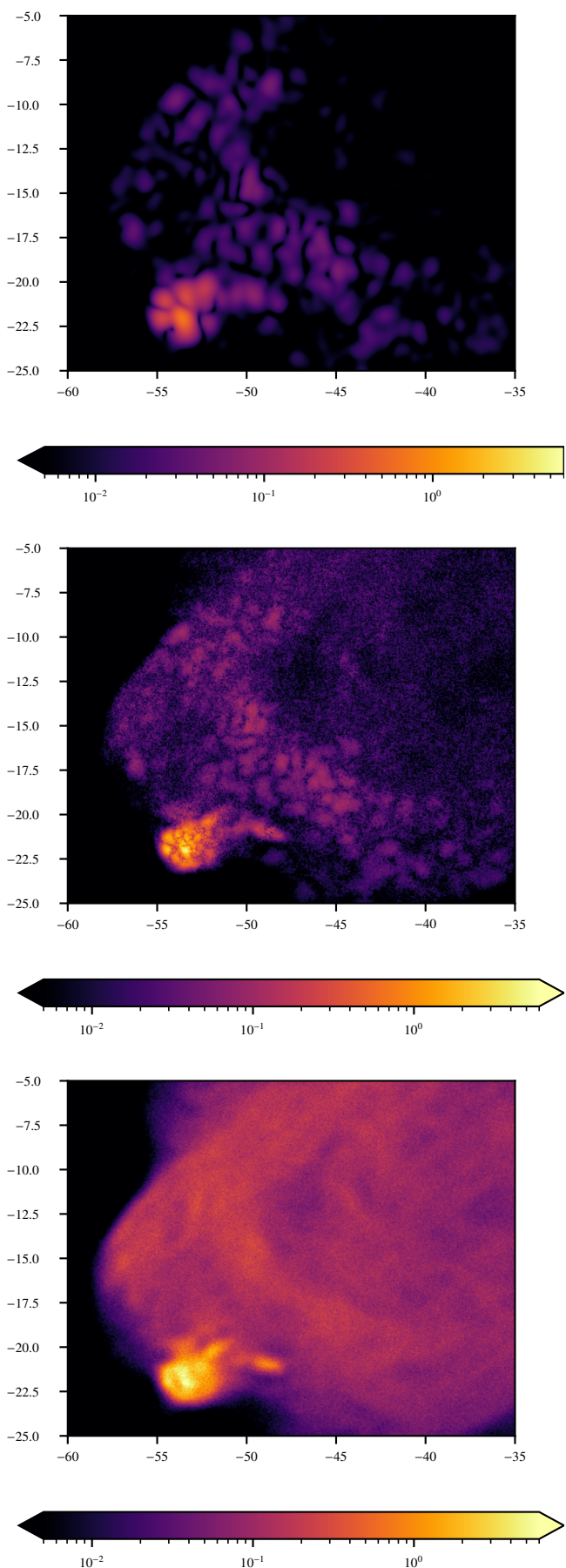}
\caption{
From left to right: linear polarization $\sqrt{Q^{2} + U^{2}}$ for \clean\, $\sqrt{\langle Q \rangle^{2} + \langle U \rangle^{2}}$ for \resolve{} and $\left\langle \sqrt{Q^{2} + U^{2}}\right\rangle$ for \resolve{} in \unit{\jy\per\as\squared}.
}\label{fig:depolarization}
\end{figure}

As a final feature of our \resolve{} polarization reconstructions, we analyze the map of linear polarization, given by $\sqrt{Q^{2} + U^{2}}$.
Exemplary, we choose the Eastern lobe for this.
In \cref{fig:depolarization}, the first row presents the linear polarization as derived from the \clean{} algorithm.
The second row depicts the linear polarization computed from the posterior means of the Stokes components via the \resolve{} algorithm.
Specifically, it is computed as $\sqrt{\langle Q \rangle^{2} + \langle U \rangle^{2}}$, a methodology that, from a Bayesian perspective, is considered incorrect.
Here and in the following, $\langle \ldots \rangle := \langle \ldots \rangle_{(I,Q,U,V|d)}$.
The third row captures the linear polarization as calculated correctly through the \resolve{} algorithm.
That is, by averaging the posterior samples of the linear polarization map: $\left\langle \sqrt{Q^{2} + U^{2}}\right\rangle$.
Upon examination, a stark difference across all three images is apparent.
Nevertheless, a commonality can be extracted from the first and second images, as both exhibit characteristic zero crossings, giving them a common visual narrative.

This narrative sheds light on the mystery of some of the observed depolarization canals found at lower frequencies in Faraday rotated radio synchrotron emission.
Such are found in low frequency polarization maps of the Milky Way and other galaxies as well as in the lobes of radio galaxies, as reported here for Cygnus A.
In the case of galaxies, the common interpretation of the depolarization canals is the combination of sharp Faraday rotation features of the intermixed synchrotron emission and Faraday rotation with \emph{beam depolarization} \citep{Shukurov2003,Haverkorn2004,Haverkorn2004a,Haverkorn2004b,Fletcher2006,Reich2006,Fletcher2007,Brentjens2011,Turic2021}.
For radio lobes, the Faraday rotation is largely external to the synchrotron emitting volume \citep{Laing1988,Garrington1988,Ensslin2003}, but could still imprint complex polarization patterns that in combination lead to depolarization canals.
Here, we argue that at least in the case of Cygnus A, but possibly also for many other radio galaxies, the observed depolarization canals have a slightly different origin than simple beam depolarization, namely \emph{uncertainty depolarization}, as we detail below.

The depolarization of initially strongly linearly polarized synchrotron emission of extended sources like the lobes of radio galaxies is in most cases due to mixing of different linear polarization states.\footnote{For compact sources, Faraday conversion to circular polarization can happen as well, but as this requires strong magnetic fields, is a very weak effect for the lobes of radio galaxies.
  And one could argue that this conversion is due to phase shifted mixing of two linear polarization components.
}
The mixing can be due to differently polarized emissions along a line of sight, within the beam area of an instrument, within a jointly imaged frequency window, or due to dislocating polarized brightness during the image reconstruction.
The reason why depolarization is more often found at low frequencies is that Faraday rotation of the polarization plane of emission traveling trough a magneto-ionized medium adds complicated structures to the received polarization pattern emitted at different 3D locations and frequencies.
This makes it easier for any of the mentioned mixing processes to erase polarized emission.

What does this imply for the fact that the \clean{} polarization and the \resolve{} $\sqrt{\langle Q\rangle^{2}+\langle U\rangle^{2}}$ maps show strong and similar, co-located depolarization canals, but the \resolve{} $\langle\sqrt{Q^{2}+U^{2}}\rangle$ map does not?
This observation suggests that a similar kind of averaging mixes polarized emission in the first two maps, but not in the latter.
The first two use point estimates of $Q$ and $U$ maps in order to construct a map of $P_\text{lin}=\sqrt{Q^{2}+U^{2}}$, whereas the latter first constructs $P_\text{lin}$ maps for each posterior sample individually, and only then averages those $P_\text{lin}$ maps afterwards.
While it is hard to understand what \clean{} does in the image construction from an information theoretical perspective, the meaning of the $\langle Q\rangle$ and $\langle U\rangle$ estimates from \resolve{} is clear; they are posterior averages of $Q$ and $U$, respectively.
The statements $\langle Q\rangle=0$ and $\langle U\rangle=0$ do not mean that necessarily $Q=0$ or $U=0$, they just mean that there is no clear indication in the data whether $Q$ or $U$ are positive, negative, or zero.

The presence of depolarization canals in the \resolve{} $\sqrt{\langle Q\rangle^{2}+\langle U\rangle^{2}}$ map therefore does not imply $Q=0$ and $U=0$  for these locations. Actually, in most posterior samples they exhibit non-zero values, making the statement that \resolve{} expects linear polarization there, but is very unsure about its direction.
Given that interferometric measurements do not probe the spatial structure directly, but through measuring individual Fourier modes, it is not always clear where a measured polarized brightness structure needs to be located on a map.
This uncertainty is expressed by $\langle Q\rangle=\langle U\rangle=0$ while simultaneously $\left\langle\sqrt{Q^{2}+U^{2}}\right\rangle>0$ holds.
This seems to be the case for all the depolarization canal locations shown in \cref{fig:depolarization}.
Thus, it might very well be that the \clean{} algorithm produces $Q$ and $U$ maps that are more close to the $\langle Q\rangle$ and $\langle U\rangle$ maps of \resolve{}.

If this interpretation is correct, it might turn out that many depolarization canals reported in literature for radio galaxies could be a consequence of \clean{} only being able to provide a point estimate of $Q$ and $U$, which -- and this would actually be a good property of \clean{} -- resembles the posterior means $\langle Q\rangle$ and $\langle U\rangle$ that \resolve{} provides.

\begin{figure*}
\centering
\loadplotpng{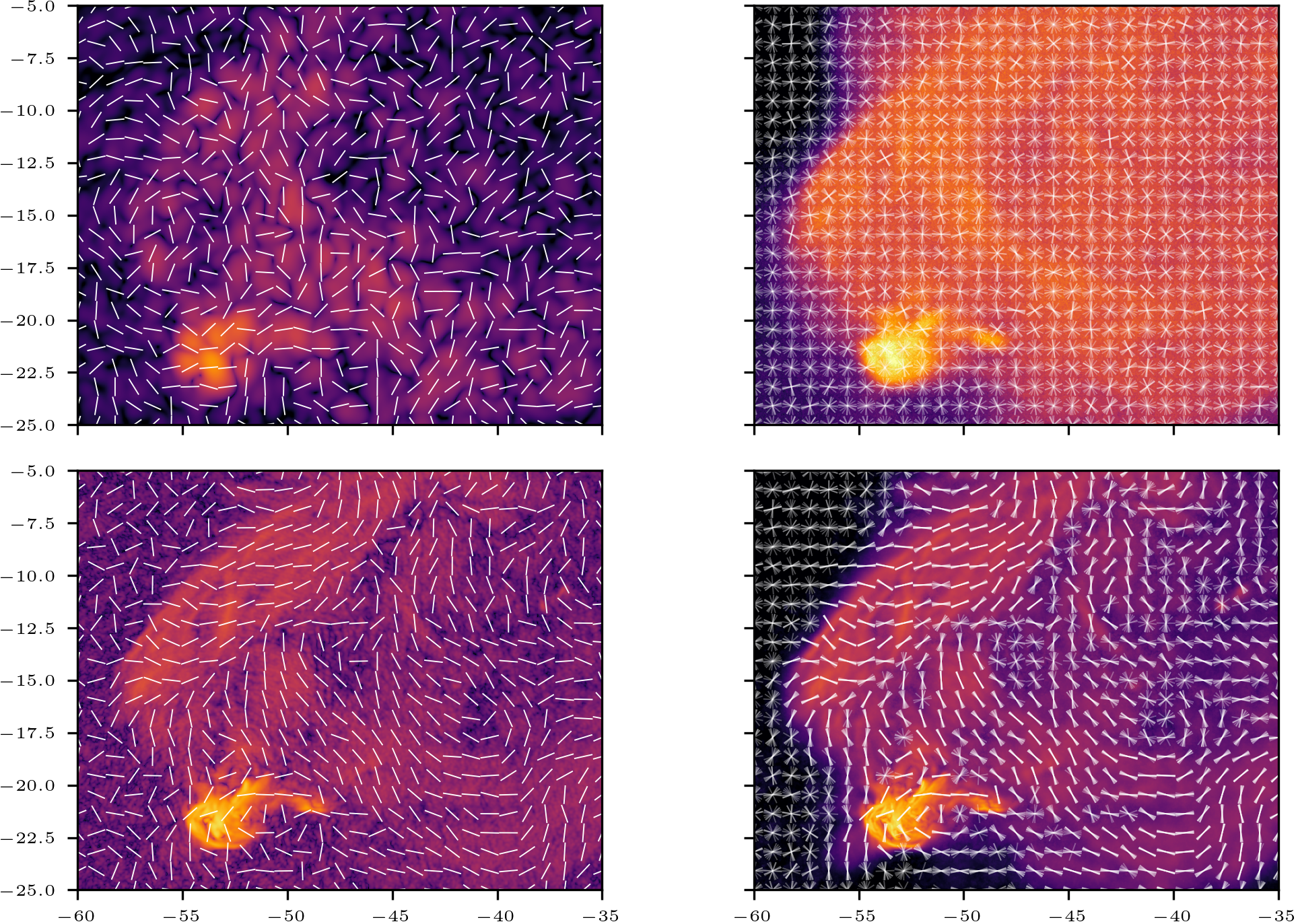}
\caption{
From left to right: \clean{} and \resolve{}. From top to bottom: 2~GHz, 13~GHz.
The background images show the linear polarized emission.
The white lines show the polarization orientation.
For \resolve{} for every available posterior sample the polarization direction is overplotted.
}\label{fig:polangle}
\end{figure*}

This interpretation is actually supported by the detailed polarization angle maps shown in \cref{fig:polangle}.
There, polarization angles for the same regions are shown at 2 and 13~GHz, for the point estimate made by \clean{} and the set of posterior samples provided by \resolve{}.
At the higher frequency of 13~GHz, there is very good agreement between the \clean{} polarization directions and those of all \resolve{} posterior samples for higher polarized brightness areas.
At lower polarized brightness areas, the \resolve{} posterior samples show more dispersion and therefore indicate an increased posterior uncertainty.
At the lower frequency of 2~GHz, where Faraday rotation is known to imprint onto the polarization structure, large polarization direction uncertainties are found for all locations of the \resolve{} map, even the one where the posterior samples agree that a larger amount of polarized brightness should exist.
\clean{} is forced to give a definite answer for $Q$ and $U$ for each location, and seems to choose $Q=U=0$ for locations \resolve{} makes the weaker statements $\langle Q\rangle=\langle U\rangle=0$.

A better understanding of the information theoretical meaning of \clean{} maps would be desirable, given the large legacy of such maps.
In that respect it is interesting to note that some recent attempt to correct biases in \clean{} polarization maps by exploiting filtered polarization direction information led to smoother depolarization canal structure \citep{Mueller2017}, indicating that at least some of the depolarization canals are rather a by-product of the used imaging method than a real property of the received polarized radiation field.
Anyhow, for future analysis, the case of depolarization should provide good arguments for proper probabilistic characterization of the posterior uncertainties, as for example done by \resolve{}.
That should prevent the occurrence of the \emph{uncertainty depolarization effect} we reported here.

The cross-like structures featured in the top right panel of \cref{fig:polangle} are due to non-isotropic uncertainties in the polarisation parameters.
For example, a larger uncertainty in $q$ and therefore $Q$ than in $u$ and $U$, respectively, leads to isotropy-imbalanced uncertainty fluctuations that appear as $+$-like patterns.
A large $q$ uncertainty means that either the $x$-direction electric field in $x$-direction or in $y$-direction is strong but not both at the same time, therefore the $+$-like pattern.
Similarly, larger $u$ (and $U$) uncertainties compared to the $q$ (and $Q$) ones create $\times$-like patterns, which are just rotated by $45^\circ$ to the former.
Other rotation angles appear due to non-vanishing uncertainty cross correlations between $q$ and $u$.

\subsection{Comparison to previous \resolve{} reconstructions}

\begin{figure*}
\centering
\input{auto/resolve_comp.pgf}
\caption{Left to right and top to bottom:
$m_0$, $m_1$, $\sigma_1 / \sigma_0$ and $(m_0 - m_1) / (\sqrt{\sigma_0^2 + \sigma_1^2})$,
where $m_0, m_1$ are the previous \citep{arras2021comparison} and the new posterior mean, respectively, and $\sigma_0, \sigma_1$ are the previous and new \resolve{} posterior standard deviation, respectively, all in \unit{\jy\per\as\squared}.
The mean within the marked box is \protect\input{auto/resolve_comp_fraction} which means that the new \resolve{} reconstruction of Stokes~$I$ has approximately half the uncertainty of the previous one.
}\label{fig:resolvecomparisonOldNew}
\end{figure*}
Until now, our focus has predominantly been on the comparative analysis of \clean{} and \resolve{} reconstructions.
We now transition to the comparison of two \resolve{} reconstructions: the pure Stokes~$I$ and the full polarization reconstruction.

Our polarisation model inherently couples all Stokes components a priori; hence, there is a flow of information between Stokes components during reconstruction.
The parameter $s$, which largely determines the strength of Stokes $I$, influences Stokes~$Q$, $U$, and $V$, and the polarization parameter $p=\sqrt{q^2+u^2+v^2}$ also influences Stokes~$I$.
This has to be the case to ensure that fractional polarization is always below 100\%, or equivalently that $P<I$.
As a result, the measured Stokes~$Q$, $U$, and $V$ signals influence the Stokes~$I$ image.
Therefore, we anticipate that the Stokes~$I$ image of the polarization reconstruction, henceforth referred to as $m_{1}$, will in some regard surpass the image reconstructed exclusively from Stokes~$I$, henceforth referred to as $m_{0}$.

\Cref{fig:resolvecomparisonOldNew} illustrates the comparison.
In the first row, $m_{0}$ and $m_{1}$ are presented.
Visual inspection suggests an increased dynamic range of $m_{1}$, although apart from this distinction the images appear congruent.

To formalize this assertion, we calculate the disparity between the images and normalize it by the uncertainty: $(m_0 - m_1) / (\sqrt{\sigma_0^2 + \sigma_1^2})$.
Observably, in regions of relatively high flux, the reconstructions align, whereas in the absence of flux, the full polarization reconstruction markedly diminishes.
This reinforces the earlier assertion of an augmented dynamic range.

Furthermore, we anticipate the uncertainty of the full polarization reconstruction to be lower, owing to the fact that the Stokes~$I$ map encompasses more information.
Indeed, when we examine the ratio of uncertainties between the two reconstructions, $\sigma_1 / \sigma_0$, we discover that the uncertainties of the full polarization reconstruction are roughly half those of the Stokes~$I$-only reconstruction in regions with substantial flux.
Notably, the polarization reconstruction is exceedingly certain about the exclusion of brightness in the low-brightness regions.

In summary, the inclusion of comprehensive polarization information not only enhances our knowledge but also improves our certainty of the Stokes~$I$ image when employing the \resolve{} algorithm.
This represents another substantial improvement over the \clean{} approach.

\subsection{Point sources}\label{sec:point-sorces}

\begin{figure}
\centering
\input{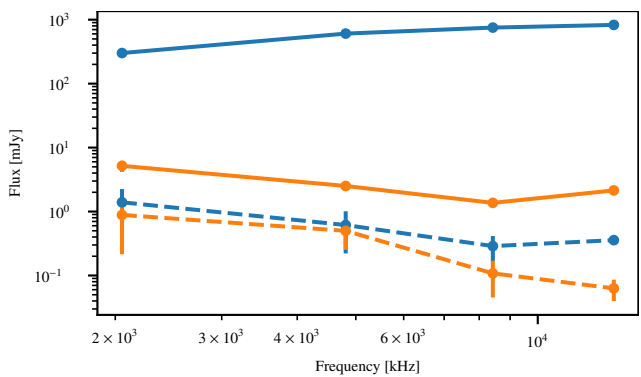}
\caption{
Point source fluxes as estimated by polarization \resolve{}.
See \cref{sec:point-sorces} for a discussion.
Blue lines: central point source (source~0), orange lines: secondary point source (source~1).
Continuous lines: Stokes I flux, dashed lines: linear polarization.
}\label{fig:pointsourceflux}
\end{figure}

The reconstructed point source flux $I^\text{p}$ of the polarization reconstruction displayed in \cref{tab:pointsourceflux} and \cref{fig:pointsourceflux} is, however, not consistent with the one found in previous works. At this stage of the development we suspect that this is more likely a result of imperfect ionospheric calibration in combination with Bayesian reconstruction than reflecting reality.
Calibration imperfections lead to inconsistent locations of point sources, to which our current model cannot adapt to.
Having now more observational constraints on the point source locations from Stokes~$I$, $Q$, and $U$ measurements increases the level of point source inconsistency and therefore lets the algorithm assign less flux to them.
Diffuse flux is much less affected by the requirement of precise co-location of different measurements and that is why we see consistency here with previous results.

\begin{table}
\begin{tabular}{lrr}
\toprule
Freq [GHz] &Source 0 [mJy] & Source 1 [mJy] \\
\midrule
\emph{Stokes I}\\
2.05&$300.7240 \pm 3.0275$*&$5.1838 \pm 1.0290$*\\
4.81&$605.8833 \pm 0.8349$*&$2.5103 \pm 0.3336$*\\
8.43&$752.3556 \pm 0.0639$*&$1.3666 \pm 0.0407$*\\
13.4&$828.2555 \pm 0.0423$*&$2.1393 \pm 0.0405$*\\
\midrule
\emph{Stokes Q}\\
2.05&$-0.1434 \pm 1.5312$&$-0.2875 \pm 0.4752$\\
4.81&$-0.0419 \pm 0.7282$&$-0.0079 \pm 0.3014$\\
8.43&$0.2072 \pm 0.1425$&$0.0371 \pm 0.0834$\\
13.4&$0.2100 \pm 0.0199$*&$-0.0376 \pm 0.0353$\\
\midrule
\emph{Stokes U}\\
2.05&$0.0253 \pm 0.7201$&$0.1099 \pm 0.9953$\\
4.81&$0.1310 \pm 0.1572$&$0.1781 \pm 0.4600$\\
8.43&$0.1696 \pm 0.0896$&$-0.0470 \pm 0.0772$\\
13.4&$0.2882 \pm 0.0262$*&$0.0143 \pm 0.0435$\\
\midrule
\emph{Stokes V}\\
2.05&$0.8536 \pm 0.2810$*&$0.0817 \pm 0.3971$\\
4.81&$0.7519 \pm 0.1198$*&$0.1646 \pm 0.1680$\\
8.43&$1.2507 \pm 0.0758$*&$-0.0430 \pm 0.0858$\\
13.4&$-0.1074 \pm 0.1526$&$0.0702 \pm 0.0427$\\
\bottomrule
\end{tabular}

\caption{
Posterior statistics for the two central point sources.
Point source~0 refers to the primary AGN and point source 1 refers to the secondary one \citep{perleytransient}.
All values that are more than $3\sigma$ away from zero are labelled with \enquote{*}.
See \cref{sec:point-sorces} for a discussion.
}\label{tab:pointsourceflux}
\end{table}

\subsection{Summary}

The application of the polarization model in \resolve{} demonstrated the superiority of polarisation-\resolve{} over \clean{} in accurately mapping polarized emission in terms of a lower level of artefacts, no unphysical negative intensity or fractional polarisation above 100\%, a higher resolution, and dynamical range.
Furthermore, some of the depolarisation canals in \clean{} maps could be explained by an incorrect averaging of polarised flux from a Bayesian perspective.
A comparison of the Stokes~$I$ maps of polarisation-\resolve{} and \resolve{} only operating on Stokes~$I$ data also shows that the former benefits from the indirect information provided by  Stokes~$Q$ and $U$ on Stokes~$I$ for the diffuse emission.
For point sources, calibration errors seem to affect the brightness reported by   polarisation-\resolve{} more severely than that by \resolve{} only operating on Stokes~$I$ data.

\section{Conclusion}\label{sec:conclusion}

In conclusion, the undertaking of polarization imaging in radio interferometry presents robust challenges, rooted in the inherent linkage and consistency constraints among the Stokes~$I$, $Q$, $U$, and $V$ maps, postulated by the principles of electromagnetism.
While this causes complexities in imaging, it also offers the opportunity to propagate information between the different Stokes parameters, yielding enhancements in image quality.
The Bayesian forward modeling approach proposed herein contrasts to the traditional backward-modeling \clean{} methodology.
Our Bayesian technique \resolve{} enables the integration of consistency constraints into the imaging process.
It offers significant advantages, including reduced noise in the Stokes~$Q$, $U$ and $V$ maps, consistent provision of uncertainty information for all Stokes components, and the ability to propagate uncertainties into derived quantities such as the polarisation angle.
Notwithstanding this, the application is computationally demanding, but remains fast enough to process VLA data.

The consistent treatment of uncertainties permitted to reveal the nature of some of the depolarization canals observed in \clean{} images.
As regions exist in polarization maps for which there must be polarized brightness but it is unclear in which direction it points, \clean{} can only give a point estimate and assigns zero flux, leading to depolarized locations.
\resolve{} is able to correctly report the presence of polarized flux, without the need to give a definite answer about its direction.
Thus, in \resolve{} maps depolarization canals of this kind are absent.

Future developments of \resolve{} include improving the prior assumptions and hence the image quality by introducing the frequency axis, speeding up the algorithm to handle larger data sets, and integrating polarisation imaging with Bayesian polarisation calibration to propagate polarisation uncertainties into the final images.
In summary, this work represents a further step towards a comprehensive Bayesian data processing algorithm for radio interferometry.

\section*{Acknowledgements}
The authors would like to thank Adrian Hamers for providing his computational resources for our reconstructions, Vishal Johnson for organizing the zenodo archive for us and Gordian Edenhofer for comments on the manuscript.
P.~Arras and J.~Roth acknowledge financial support by the German Federal Ministry of Education and Research (BMBF) under grant 05A20W01 (Verbundprojekt D-MeerKAT) and 05A23WO1 (Verbundprojekt D-MeerKAT).

\bibliographystyle{aa}
\bibliography{bib.bib}
\end{document}